# RECTANGULAR POTENTIAL BARRIER AFFECTED BY EXTERNAL FIELDS, HYDROSTATIC PRESSURE AND IMPURITIES

[1]*Julián A. Zúñiga*, [2]*Ober L. Hernández R*, [2]*S. T. Pérez-Merchancano*.

**1.** *Departamento de Matemáticas, Universidad del Cauca calle 5 # 4-70*

**2.** *Departamento de física, Universidad del Cauca calle 5 # 4-70*

*Popayán- Cauca Colombia.*

In this work the influence of the electric and magnetic fields over a tunneling particle in a rectangular potential barrier is shown, we have taken into account the presence of an impurity at the barrier center and the effects of a hydrostatic pressure parallel to the barrier height considering the BenDaniel-Duke boundary conditions. Given that the particle is moving inside a GaAs-Al$_x$Ga$_{1-x}$As-GaAs heterostructures it is evident a change in the transmission coefficient due to the impurity concentration and the presence of the hydrostatic pressure. The potential due to the presence of the impurity is approximate with a second degree polynomial function that resolves the discontinuity generated by heavily modifying the transmission coefficient.

1. **Introduction**

The low dimensional GaAs-Al$_x$Ga$_{1-x}$As-GaAs heterostructures has been widely accepted for the analysis and construction of ultrafast devices, this is possible, on one side, to the better understanding of the growing kinetics of types III–V materials in comparison to other semiconductor compounds, on the other side, for any practical application it is necessary for the semiconductor to have a great band discontinuity, high mobility and that its growth with type p and n impurities could be easily controlled. In this case we have taken into account the BenDaniel-Duke boundary conditions [1].

2. **Methodology**

The system is shown in the figure 1 where the barrier height $\upsilon_0(p,\chi)$ depends of the hydrostatic pressure (hp) and the material concentration, the particle is moving along the z axis, the electric field $\vec{F}$ is parallel to the tunneling direction of the particle, and the magnetic field $\vec{B}$ is perpendicular to the yz plane, besides an $z_0$ impurity has been considered that for this particular is located in the barrier center.

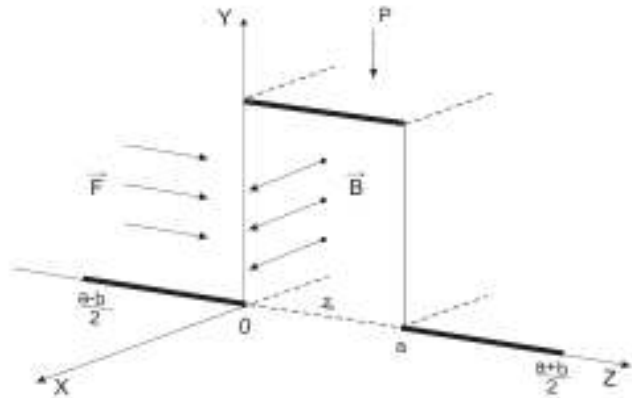

**Figure 1.** Sketch of a three-dimensional model of electron tunneling. Transmission of electrons through the potential barrier *V* of width *a* grown along z containing an impurity.

For this results we have considered the effective mass $m_I^*(p)$ and the medium permittivity constant $\varepsilon(p)$ as a function of the hp [2, 3], as is seen on the 1 and 2 equations.

$$m_I^*(p) = m_I^*(0) e^{0.0078 p} \tag{1}$$

$$\varepsilon(p) = 1.02132 * \varepsilon_0 \, e^{-1.67 \times 10^{-3} p} \tag{2}$$

The barrier potential is in terms of the hp as follows:

$$V(z,p) = \begin{cases} 0 & si \quad \dfrac{a-b}{2} < z < 0 \\ v_0(p,\chi) & si \quad 0 < z < a \\ 0 & si \quad a < z < \dfrac{a-b}{2} \end{cases} \tag{3}$$

The impurity generated potential inside the barrier is approximated as follows:

$$V_e^{(\xi)} = \dfrac{e^2}{z_0^2 \, \varepsilon(p) \xi} \begin{cases} z^2 + z_0 z + z_0^2 & si \quad 0 < z < z_0 \\ z^2 - 5 z_0 z + 7 z_0^2 & si \quad z_0 < z < a \end{cases} \quad ; \quad \xi \geq \dfrac{k_B^4 \, k_e}{z_0^3} \tag{4}$$

The term $\xi$, helps in the approximation to the impurity potential.

The wave equation that involves the previously established parameters in the system heterostructure is given by:

$$\dfrac{\partial^2 \psi}{\partial z^2} - \left( \dfrac{1}{k_B^4} z^2 + \dfrac{1}{k_F^3} z + k_2^2 \right) \psi + \dfrac{k_e}{|z - z_0|} \psi + \dfrac{\partial^2 \psi}{\partial y^2} - i \dfrac{2}{k_B^2} z \dfrac{\partial \psi}{\partial y} = 0 \tag{5}$$

Where $k_F$ and $k_B$ are the electric and magnetic longitudes due to the electric field $\vec{F}$ and magnetic field $\vec{B}$, the wave function that is solution to this systems is:

$$\psi(y,z) = \begin{cases} A \, e^{i k_1 z} + B \, e^{-i k_1 z} & si \quad \dfrac{a-b}{2} < z < 0 \\ e^{-\frac{1}{2}\eta + i k_y y} \left[ C \, M\left( \dfrac{\lambda+1}{4}; \dfrac{1}{2}; \eta \right) + D \, U\left( \dfrac{\lambda+1}{4}; \dfrac{1}{2}; \eta \right) \right] & si \quad 0 < z < z_0 \\ e^{-\frac{1}{2}\gamma + i k_y y} \left[ E \, M\left( \dfrac{\lambda+1}{4}; \dfrac{1}{2}; \gamma \right) + F \, U\left( \dfrac{\lambda+1}{4}; \dfrac{1}{2}; \gamma \right) \right] & si \quad z_0 < z < a \\ G \, e^{i k_1 z} & si \quad a < z < \dfrac{a+b}{2} \end{cases} \tag{6}$$

Where the arguments of the hypergeometric confluent functions involve the external fields, the impurity effects and the hp. Therefore, the particle transmission coefficient is given by the expression:

$$T = \dfrac{1}{R^2} e^{(\eta_0 + \eta_{z_o} + \gamma_{z_o} + \gamma_a)} \left[ (w_1 + w_3)^2 + (w_2 + w_4)^2 \right]^{-1} \tag{7}$$

## 3. Results and Conclusions

When finding the transmission coefficient, the wave amplitudes C, D, E and F contain the $e^{-ik_y y}$ factor, therefore the wave function depends only on the tunneling direction of the particle, as seen in figure 1. The existing relationship between the hp and the concentration in the tunneling is seen on figure 2 where it is noticeable that at a bigger concentration the tunneling probabilities increases and the hp effects are more evident as its increments from 0 to 30 Kbar. Also in the figure 2 there is a lower peaks concentration suggesting a lower tunneling, this is confirmed by the figure 3 where the hp is seen not favoring the tunnel effect. It is relevant to note that in this process a 500 Kv/cm electric field and a 0.6 T magnetic field were used.

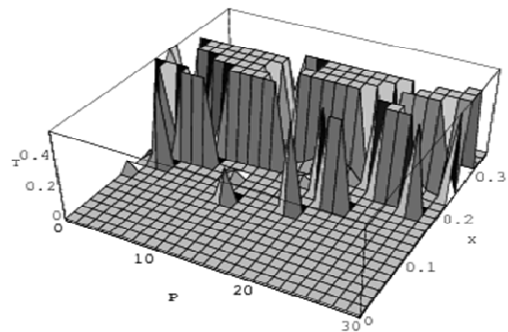

**Figure 2.** Transmission coefficients as function of material concentration and hydrostatic pressure.

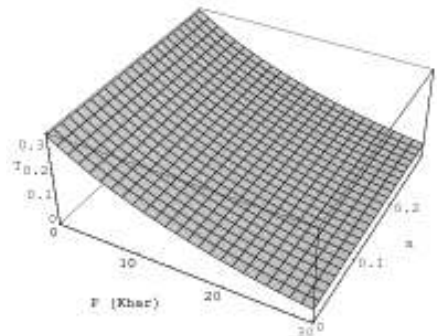

**Figure 3.** Transmission coefficients as a function of the hydrostatic pressure.